\documentclass[a4paper,conference]{IEEEtran}

\IEEEoverridecommandlockouts % To enable \thanks command

\usepackage{tikz}
\usepackage{textcomp}
\usepackage{hyperref}
\usepackage{lipsum}

\newcommand\copyrighttext{%
  \footnotesize \textcopyright 2024 IEEE. Personal use is permitted, but republication/redistribution requires IEEE permission. This article has been accepted for publication in 2024 27th Conference on Innovation in Clouds, Internet and Networks (ICIN).  Citation information: DOI \href{https://doi.org/10.1109/ICIN60470.2024.10494463}{10.1109/ICIN60470.2024.10494463}}
\newcommand\copyrightnotice{%
\begin{tikzpicture}[remember picture,overlay]
\node[anchor=south,yshift=800pt] at (current page.south) {\fbox{\parbox{\dimexpr\textwidth-\fboxsep-\fboxrule\relax}{\copyrighttext}}};
\end{tikzpicture}%
}

\usepackage{graphicx} % Required for inserting images
\usepackage{amsmath}
\usepackage{amssymb}
\usepackage{amsfonts}
\usepackage{subcaption}
\usepackage{algorithm}
\usepackage{algpseudocode}
\usepackage{booktabs}
\usepackage{url}
\usepackage{hyperref}
\usepackage{setspace}

\title{Cost Minimization in Multi-cloud Systems with Runtime Microservice Re-orchestration}%\thanks{~\copyright~2024IEEE. Personal use of this material is permitted. Permission from IEEE must be obtained for all other uses, in any current or future media, including reprinting/republishing this material for advertising or promotional purposes, creating new collective works, for resale or redistribution to servers or lists, or reuse of any copyrighted component of this work in other works.}}
\author{Marco Zambianco, Silvio Cretti, Domenico Siracusa}
\author{
    \IEEEauthorblockN{Marco~Zambianco\IEEEauthorrefmark{1}, Silvio Cretti\IEEEauthorrefmark{1}, Domenico Siracusa\IEEEauthorrefmark{1}}
        \IEEEauthorblockA{\IEEEauthorrefmark{1}Fondazione Bruno Kessler, Trento, Italy
\\Email: mzambianco, scretti, dsiracusa@fbk.eu}       
}

\date{October 2023}

\begin{document}

\maketitle

\copyrightnotice

\begin{abstract}
Multi-cloud systems facilitate a cost-efficient and geographically-distributed deployment of microservice-based applications by temporary leasing virtual nodes with diverse pricing models. To  preserve the cost-efficiency of multi-cloud deployments, it is essential to redeploy microservices onto the available nodes according to a dynamic resource configuration, which is often performed to better accommodate workload variations.
However, this approach leads to frequent service disruption since applications are continuously shutdown and redeployed in order to apply the new resource assignment. To overcome this issue, we propose a re-orchestration scheme that migrates microservice at runtime based on a rolling update scheduling logic. Specifically, we propose an integer linear optimization problem that minimizes the cost associated to multi-cloud virtual nodes and that ensures that delay-sensitive microservices are co-located on the same regional cluster. The resulting rescheduling order guarantees no service disruption by repacking microservices between the available nodes without the need to turn off the outdated microservice instance before redeploying the updated version. In addition, we propose a two-step heuristic scheme that effectively approximates the optimal solution at the expense of close-to-zero service disruption and QoS violation probability. Results show that proposed schemes achieve better performance in terms of cost mitigation, low service disruption and low QoS violation probability compared  to baseline schemes replicating Kubernetes scheduler functionalities.

\end{abstract}
\section{Introduction}
Multi-cloud systems enables the flexibility to lease virtual nodes from various cloud owners. By combining resources from multiple providers that offer different pricing models, organizations can implement cost-effective optimization strategies for the deployment of  their microservice-based applications. \cite{hong2019overview}. 

In a multi-cloud environment, from a service provider standpoint, the resource utilization efficiency of the deployment plays a central role to minimize the amount of rented multi-cloud virtual nodes required to run the various microservices composing each application \cite{georgios2021exploring}. To ease the management and scalability of the deployment,  container orchestration frameworks such as Kubernetes are increasingly used to automatize the allocation of microservices on the available nodes based on their resource requirements \cite{vayghan2018deploying}. 
 However, the geographical distribution of the underlying multi-cloud physical infrastructure increases the complexity of the cost minimization objective using these tools. As a matter of fact, resource fragmentation and high network latency across nodes in different regions introduce additional challenges into the microservice scheduling process, as they can affect the dependability of the Quality of Service (QoS) performance of the applications \cite{tamiru2021mck8s}.

To overcome this challenge, the research activity has focused on the design of custom orchestration schemes to augment state-of-the-art orchestrators functionalities and compute cost-efficient deployment configurations of the various applications while, at the same time, fulfilling the related QoS requirements \cite{senjab2023survey}. In detail, these solutions provide a fixed deployment configuration in which microservices are assigned to specific virtual nodes for the whole duration of their life-cycle according to some cost model and QoS criterion.
%However, microservices may require a rescaling of the allocated resources due to several reasons including the adjustment of task completion time based on workload intensity regime, handling seasonal user requests peaks, supporting new features rollout, improving the cluster resource utilization efficiency 
However, microservices often necessitate a reallocation of the assigned resources for various reasons. These reasons may include adjusting task completion times in response to workload fluctuations, managing seasonal spikes in user requests, facilitating the introduction of new features, mitigating resource contention on overloaded clusters  \cite{baarzi2021showar}.

%which in turn affect their resource efficiency performance and thus the cost efficiency performance. 

As a consequence, a periodical deployment re-orchestration (in other words, the rescheduling of already-deployed microservices) can further improve cost savings by tailoring the deployment configuration to the updated resource request \cite{mendoncca2019developing}. %As a results, scaling strategies, such as vertical and/or horizontal scaling, are frequently applied to adjust the resource assigned to each microservices and prevent overload/underload scenarios. The continuous rescaling of the assigned resources deteriorates the cost-effectiveness of the initial deployment configuration as resized microservices progressively be rescheduled on different nodes. Consequently, periodical deployment re-orchestration (in other words, the rescheduling of already-deployed microservices) can improve cost savings for service providers by repacking microservices leveraging multi-cloud architectures by tailoring the number of multi-cloud virtual nodes according to the actual resource request. 
Such re-optimization must be carried out at runtime in order to prevent any service disruption caused by recreating the whole application deployment (in other words, by turning off current microservice instances before redeploying the new instances according to the updated resource configuration). In this context, existing cost-aware orchestration schemes are unpractical as they assume that microservices are scheduled for the first time, thereby implicitly requiring the recreation of the whole deployment at each new re-orchestration phase to take advantage of possible cost reductions.% allocation assuming  implicitly assume the deployment recreation assume that microservices are scheduled once and that are hosted on the same virtual nodes for all the duration of its life-cycle, are unpractical since they would recreatunsuited since they ignores any disruption-free microservice re-orchestration mechanisms, since the allocation of each microservice is predetermined and remains unchanged during its life-cycle. Consequently, they provide largely suboptimal results 
 
 %In particular, the shut down of outdated microservice instances and their subsequent re-deployment with an updated resource request configuration would cause a non-negligible service disruption. Consequently, existing schemes, that do not include microservice re-orchestration mechanisms,  have limited efficacy since the potential cost savings gained by a new re-deployemt of the applications is likely to be out-weighted by the QoS performance degradation.
Alternatively, a deployment re-orchestration performed as a rolling update, that progressively deploys microservices according to the new resource request and subsequently terminates the old instances,  ensures a disruption-free re-scheduling procedure \cite{singh2017container}. %with close-to-zero impact on the QoS performance. 
To fully benefit from this deployment update technique, it is essential to account for the simultaneous coexistence of the new and old microservice instances, temporarily increasing the resource occupation of the deployment, as well as the variable resource load on each virtual node that continuously increases/decreases during the rolling update procedure. These dynamics largely affects the optimality of the resulting deployment in terms of cost minimization and QoS fulfillment if they are not handled by a suitable re-orchestration algorithm specifically designed according to a rolling update logic.

Following these observations, we propose a disruption-free re-orchestration algorithm that, given an initial deployment configuration, reschedules microservices in a rolling-update fashion to minimize the deployment cost while preserving the QoS performance. In detail, we summarize our contributions as follows:
\begin{itemize}
    \item We design an integer linear programming problem, based on a customized version of the Bin Packing Problem,  that minimizes the deployment cost by repacking microservices  on a suitable number of multi-cloud virtual nodes  while preserving QoS  performance. In particular, we analytically express a constraint on the microservice rescheduling order such that each microservice can be migrated at run-time without the need to first turn off its outdated instance.
    \item To overcome the computational complexity of the proposed formulation, we design a two-phase heuristic algorithm that provides comparable performance in terms of deployment cost as the optimal solution while ensuring a low QoS violation probability and limited service disruption. %the optimal solution by first computing the new virtual node assignment for each microservice and then computing a suitable reording of each rescheduling step. The proposed scheme computes the rescheduling of microservice by first computing a new microservice allocation on the available virtual nodes in order minimizes the dpeloyment cost and preservce the QoS. Then by computing a suitable microservice rescheduling order that migrates microsreeach task of the optimal solution (i.e., and low disruption ) separately in order to provide practical solution for real-world scenarios.
    \item We assess the performance of proposed optimal and heuristic re-orchestration schemes against baselines, replicating the Kubernetes scheduling behavior, in terms of cost minimization, disruption cost and QoS violation probability performance. The results show that our solutions consistently outperform the considered benchmarks across all metrics. 
\end{itemize}

The remainder of the paper is organized as follows. In Section II, we analyze the state of the art. In Section III, we describe the considered system model. In Section IV, we present the proposed re-orchestration scheme. In Section V, we discuss the obtained results. Finally, we draw the conclusion in Section VI.

\section{Related work}
%The advent of multi-cloud computing has increased the complexity of orchestration schemes due to the distributed nature of such system. In such scenario, many works have addressed the problem of deployment cost minimization by means of a combination of microservice scheduling and cluster autoscaling strategies. 
An overview of the main challenges and proposed solutions in the context of multi-cloud resource orchestration can be found in \cite{tomarchio2020cloud}.  The authors of \cite{jiang2020cloud} design an orchestration framework that enables a cost-efficient scaling of data-intensive applications across multiple geo-distributed clusters. Similarly, \cite{wang2020elastic} proposes an elastic scheduling scheme  that combines microservice scaling and allocation to optimally minimize the number of required virtual nodes according to their cost. Although these works offer resource-efficient orchestration schemes, they are designed to compute the initial deployment configuration, thus their optimality is uncovered when applied to an already-running deployment. Conversely, we design a re-orchestration scheme that re-optimizes the cost of already deployed microservices by migrating them across the available nodes. The authors of \cite{aldwyan2021elastic} propose a microservice scheduling framework, based on a genetic algorithm, to relocate microservice on multi-cloud clusters in order to minimize the deployment cost and service latency. However, the relocation solution is performed by recreating the deployment, thus introducing service disruption. Differently, we design a disruption-free rescheduling algorithm implementing a rolling update scheduling logic. 

The authors of \cite{zhong2020cost} and \cite{sampaio2019improving}  propose heuristic-based microservice rescheduling schemes to automatically co-locate microservices having high affinity values in terms of number of exchanged data packets in order to further improve the cost mitigation. However, their approach neglects the cost associated to the virtual nodes. Moreover, the enforcement of microservice locality constraints  cannot reliably be satisfied by affinity rules, which act as soft constraints. Differently, we first design an optimization problem that provides the optimal microservice rescheduling sequence that minimizes the number of used virtual nodes while guaranteeing the locality constraints fulfillment, then we approximate its solution with a suitable low-complexity heuristic algorithm.%according to between microservices   
%that first optimizes the initial placement exploiting already available resources, then it reschedules microservices to better accomodate workloads. Similarly, the authors of \cite{sampaio2019improving} design a runtime microservice placement scheme . D

\section{System model}

\begin{figure}[th]
    \centering
     \includegraphics[width=0.5\textwidth]{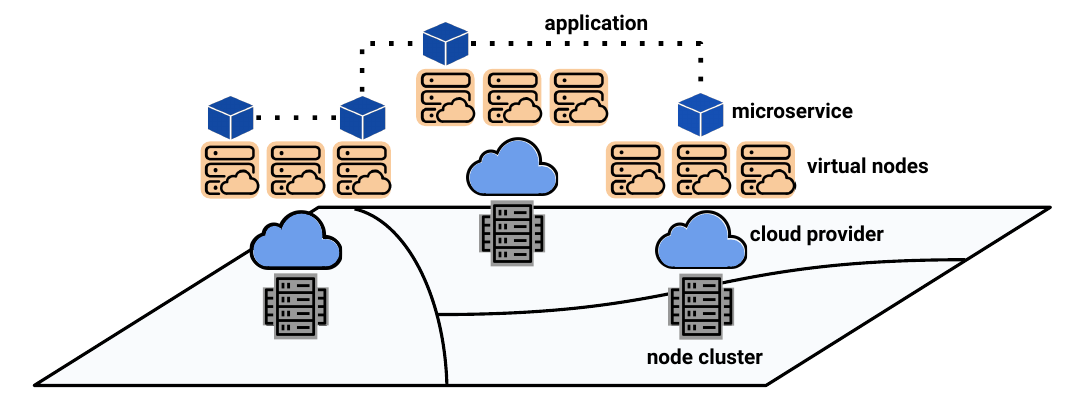}
     \caption{Multi-cloud system model. Microservices of each application are deployed on various virtual nodes belonging to different geographically-distributed cloud providers.}
     \label{fig:sys_model}
\end{figure}

We consider a service provider that deploys a variety of microservice-based applications encompassing e-commerce platforms, content streaming services, machine learning tasks. We indicate the set of deployed applications as $A$ and the set of microservices composing each application as $M_a, a \in A$. To increase the scaling capability according to the incoming workloads, the various microservices are deployed on virtual nodes belonging to $R$ geographically distributed clusters. In particular, one cluster is owned by the service provider, whereas the remaining ones are administrated by different cloud providers (e.g., AWS, Microsoft Azure, Google Cloud, etc.). We indicate the set of virtual nodes in the private cluster as $N_0$, whereas we indicate the set of virtual nodes in each multi-cloud cluster as $N_1,...,N_{R-1}$. Moreover, we comprehensively represent the set of virtual nodes as $N=\bigcup_{i=0}^{R-1} N_i$. We depict an overview of the considered multi-cloud model in Fig. \ref{fig:sys_model}. %he deploys his applications both in his on-premise private cloud, composed of a single cluster $R_0$ of $N_{R_0}$ virtual nodes, and in a multi-cloud system composed by geo-distributed $R_1,...,R_P$ clusters of $N_{R_1},..., N_{R_P}$ each that are administrated by a corresponding number of cloud providers (e.g. AWS, Microsoft Azure, Google Cloud, etc.). We comprehensively represent the set of active virtual nodes as $N = \{0,1,...,N_{R_0},...,N_{R_0}+N_{R_1},...,\sum_{i \in P} N_{R_i}\}$ (intuitively, we group virtual nodes within the same cluster and we indicate them by incrementally assigning an integer value starting from zero). 

Due to the geo-distribution of the cloud infrastructure, the network communication between virtual nodes belonging to different clusters experiences some non-negligible latency denoted as $D_{n,n'} > 0$, $n \in N_i$ , $n' \in N_{j \neq i}$. Conversely, we assume that intra-cluster network latency can be approximated as zero, thus $D_{n,n'} = 0$, $n, n' \in N_i$.
Each microservice is scheduled to demand a minimum of $r^{(cpu)}_{a,m}$ CPU and $r^{(ram)}_{a,m}$ RAM resources from the assigned virtual node that provides a maximum of $C^{(cpu)}_n$ CPU and $C^{(ram)}_n$ RAM. Note that we generally refer to the amount of resources as $r_{a,m}$ and  $C_n$ for the sake of notation clarity. We indicate the assignment of microservices on the various virtual nodes as $s_{a,m,n} = 1$ if $m \in M_a, a \in A$ is scheduled on node $n \in N$, $s_{a,m,n} = 0$ otherwise. 
Moreover, to prevent the degradation of QoS performance resulting from synchronization issues between microservices that communicates across virtual nodes in different clusters, microservices may be deployed according to some locality constraints which enforce their co-location on virtual nodes within the same cluster. In detail, the service provider imposes a maximum tolerable network latency, denoted as $d_{a,m,m'}$, on each existing communication flow between microservices $m,m' \in M_a,  \forall a \in A$. As a result, we express the location feasibility for each pair of communicating microservices belonging to the same application by defining the indicator variable $\ell^{(n,n')}_{a,m,m'} = 1$ if $d_{a,m,m'} \geq D_{n,n'} \quad  \forall m,m' \in M_a, \forall n, n' \in N$, $\ell^{(n,n')}_{a,m,m'} = 0$ otherwise. In other words, $\ell^{(n,n')}_{a,m,m'}$ indicates whether the allocation of microservices $m$ and $m'$ on virtual nodes $n$ and $n'$ complies with the latency requirements. 
Based on this notation, we analytically represent a feasible deployment configuration as %of each microservice $m \in M_a,a \in A$ on the $N$ virtual nodes as% $s_{a,m,n}, \forall a \in A, m' \in M_a, n \in R_i  $
\begin{multline}
S_A = \{ s_{a,m,n} : \sum_{a' \in A} \sum_{m' \in M_a} s_{a',m',n'} r_{a',m'} \leq C_n \quad \land  \\
    s_{a,m',n'} = 1 \quad \text{if} \quad  \ell^{(n,n')}_{a,m,m'} = 1, \forall m' \in M_a, n' \in N \}  
\end{multline}
Essentially, from the above definition, a feasible microservice deployment must not exceed the virtual nodes maximum resource capacity and, at the same time, must satisfy the locality constraints.

We assume that the service provider computes the initial deployment of its applications using some custom orchestrator scheme that leverages Kubernetes scheduling functionalities to create a feasible deployment $S_A$. The latter has associated an economic cost that depends on the amount of virtual nodes rented from the different cloud providers. In detail, we indicate the price per unit of resource associated to each virtual node as  $p_n, n \in N$, where $p_n > 0$  if $n \in N_{i}, i \in \{1,...,R-1\}$,  $p_n = 0$ otherwise (we assume that virtual nodes in the private cluster are cost-free since most expenses are related to the utilities costs associated to run the physical infrastructure). 

The service provider may vertically resize (in other words, scale up or down) the amount of resources allocated to the deployed microservices in order to achieve a resource-efficient accommodation of various workloads regimes. For example, it may temporally scale up the resources to handle user request spikes and later restoring the original resource configuration once the request burst ends. We acknowledge that horizontal scaling is the most common choice to handle resource scaling, however, besides providing a less granular resource control compared to vertical scaling and thus less preferable when the deployment cost minimization is the main optimization goal \cite{kwan2019hyscale}, it also implies a proper management of the various replicas to enhance the fault tolerance of the deployment. For this reason, we leave the analysis of a re-orchestration scheme that includes horizontal scaling as a future work.
%Such resource rescaling can be caused by several reasons such as better workload accommodation, handling seasonal user requests peaks, new feature rollout, inefficient resource utilization.

We indicate as $\hat{r}_{a,m}$ the new amount of resources requested by microservice $m \in M_a$. To reduce the deployment cost while ensuring no service disruption, the service provider employs a re-orchestration scheme to reschedule at runtime a subset $\hat{A}$ of deployed applications in order to minimize the number of required multi-cloud virtual nodes based on their price. More precisely, $\hat{A}$ includes applications whose microservices resource configuration has been updated as well as a number of non-reconfigured applications (i.e., applications whose microservices resource configuration has not been updated, hence $\hat{r}_{a,m} =  r_{a,m}$). The latter are applications that may be rescheduled in order to possibly improve the overall virtual nodes resource utilization by repacking microservices in a smaller amount of nodes, thus leading to further costs saving. 

The re-orchestration procedure is executed as follows. For each rescheduling slot $t \in T$, where $|T| = |\bigcup_{a \in \hat{A}} M_a|$, a single microservice $m \in M_a, \forall a \in \hat{A}$ is rescheduled on the available nodes according to a rolling update logic: first the updated microservice instance is allocated on the destination virtual node, then the old existing microservice instance  is terminated. 

On one hand, such redeploying procedure ensures zero service disruption since the tasks processed by each rescheduled microservice are handed over to the updated copy before terminating the old instance (note that this is not the case if the rescheduling procedure followed a recreate approach, where old microservice instances are terminated first). On the other hand, the sequential rescheduling of microservices  increases the re-orchestration  complexity. The simultaneous presence of both new and old microservice instances temporary increases the resource occupation on the virtual nodes, causing continuous fluctuations in resource load during the rescheduling procedure. As a consequence, an effective re-orchestration algorithm must compute a suitable rescheduling order handling such dynamic in order to optimally minimize the deployment cost. Furthermore, the complexity of such task is further exacerbated by the locality constraints that must be preserved after the rescheduling procedure completion. In the next section, we design an optimization problem addressing these challenges.

\section{Problem formulation}
\subsection{Optimal re-orchestration solution}
We propose an integer linear problem formulation to compute the optimal disruption-free rescheduling order that, starting from a deployment configuration $S_A$,  produces a new feasible deployment configuration $S_{\hat{A}}$ minimizing the deployment cost. 
We introduce the binary optimization variable $y_n, n \in N$ indicating if a virtual node is hosting any microservice, hence $y_n = 1$ if $\exists m \in M_a, \forall a \in A$ such that $s_{a,m,n} = 1$, $y_n = 0$ otherwise. Similarly, we denote the binary optimization variable $x^{t}_{a,m,n}$ that indicates when the allocation of a microservice on a selected node is performed, hence $x^{t}_{a,m,n} = 1$  if microservice $m \in M_a, \forall a \in \hat{A}$ is allocated on node $n \in N$ during the rescheduling time slot $t \in T$, $x^{t}_{a,m,n} = 0$ otherwise.  Moreover, we denote as $\Delta C_n = C_n - \sum_{a \in A}\sum_{m \in M} s_{a,m,n} r_{a,m}, \forall n \in N$ the available capacity on each virtual node according to the initial deployment configuration $S_A$. Leveraging these variables, we design the following optimization problem as

\begin{equation} \label{eq_obj_func}
   \min_{y,x} \sum_{n \in N} p_n C_n \cdot y_n 
\end{equation}

Subject to

\begin{equation}\label{eq_cons1}
    \sum_{a \in \hat{A}} \sum_{m \in M_a} \sum_{t \in T} \hat{r}_{a,m} x^{t}_{a,m,n} \leq y_n \Delta C_{n}  \quad \forall n \in N
\end{equation}

\begin{equation}\label{eq_cons2}
    \begin{split} \sum_{t \in T} x^{t}_{a,m,n}  + \sum_{t \in T} x^{t}_{a,m',n'} \leq 1 + \ell^{(n,n')}_{a,m,m'} \\ \forall m,m' \in M_a,  \forall a \in \hat{A}, \forall n,n' \in N \end{split}
\end{equation}

%\begin{equation}\label{eq_cons3}
%\begin{split}
%    x^{t}_{a,m,n} \hat{r}_{a,m} \leq \Delta C_n + \\  \sum_{t' \in T_{t'< t}}  \sum_{n' \in N} \Big[ \sum_{a' \in \hat{A}} \sum_{m' \in M_a} 
%                                         s_{a',m',n} \Big( x^{t'}_{a',m',n'}   r_{a,m}  \Big) \\ -
 %                                    \sum_{a' \in \hat{A}} \sum_{m'\neq m \in M_a}   s_{a',m',n'} \Big( x^{t'}_{a',m',n}   \hat{r}_{a,m} \Big) \Big]\\  
 %                                     \forall a \in \hat{A}, \forall m \in M_a,  \forall n \in N, \forall t \in T
%\end{split}
%\end{equation}

\begin{equation}\label{eq_cons3}
\begin{split}
     \sum_{t' \in T_{t'< t}} \Big{\{} \sum_{n' \in N} \Big[ \sum_{a \in \hat{A}} \sum_{ m \in M_a}   s_{a,m,n'} \Big( x^{t'}_{a,m,n}   \hat{r}_{a,m} \Big) - \\ \sum_{a' \in \hat{A}} \sum_{m' \neq m \in M_a} 
                                         s_{a',m',n} \Big( x^{t'}_{a',m',n'}   r_{a',m'}  \Big)  \Big] \Big{\}} \leq \Delta C_n
                                      \\ \quad \forall n \in N, \forall t \in T
\end{split}
\end{equation}

\begin{equation}\label{eq_cons4}
   \sum_{t \in T} \sum_{n \in N}  x^{t}_{a,m,n} = 1 \quad \forall m \in M_a, \forall a \in \hat{A}
\end{equation}

\begin{equation}\label{eq_cons5}
   \sum_{m \in M_a} \sum_{n \in N}  x^{t}_{a,m,n} = 1 \quad \forall t \in T, \forall a \in \hat{A}
\end{equation}

\begin{equation}\label{eq_cons6}
   y_n = 1 \quad \forall n :  s_{a,m,n} = 1, m \in M_a, a \in A \backslash \bar{A} \ 
\end{equation}

\begin{equation}\label{eq_cons7}
   y_n, x^{t}_{a,m,n} \in \{0,1\} \quad \forall a \in \hat{A}, \forall m \in M_a,  \forall n \in N, \forall t \in T
\end{equation}

The objective function \eqref{eq_obj_func} computes the total deployment cost as a summation of the costs associated to each multi-cloud virtual node. Constraint \eqref{eq_cons1} ensures the final deployment resource occupation is feasible by guaranteeing that the new resource configuration of the rescheduled microservices does not exceed the assigned node capacity. Moreover, it links the rescheduling optimization variable $x^{t}_{a,m,n}$ to the node optimization variable $y_n$ by making sure that unused nodes are not eligible for rescheduling. Constraint \eqref{eq_cons2} enforces locality constraints by guaranteeing that a given microservice pair $m,m'$ cannot be allocated on nodes $n$ and $n'$ in different clusters if $\ell^{(n,n')}_{a,m,m'} = 0$. Constraint \eqref{eq_cons3} implements the rescheduling procedure as rolling update by ensuring that in each time slot $t$ the rescheduled microservices in previous times slot $0,..,t',..t$ has never exceeded the available capacity $\Delta C_n$ in each node. In detail, the first addend within the square bracket computes the resource occupation of each microservice $m$ rescheduled on node $n$  according to the updated resource configuration $\hat{r}_{a,m}$ given the initial deployment configuration $s_{a,m,n}$ (note that we also account for microservices that can be rescheduled on the same node when $n'=n$, meaning that the old and updated microservice instances coexist on the same node during time slot $t'$). Conversely, the second addend computes the amount of resources $r_{a,m}$ freed up by the rescheduling of each microservice $m' \neq m$ that was initially allocated on node $n$ and has been relocated on node $n'$. In other words, this expression progressively monitors in each scheduling slot the increase/decrease of resource occupation in each virtual node caused by the rescheduling of microservices from/to other active virtual nodes. Constraints \eqref{eq_cons4} and \eqref{eq_cons5} guarantees that every microservice is rescheduled once in each time slot $t$ and that it is uniquely assigned to node $n$, respectively. Constraint \eqref{eq_cons6} makes sure that the virtual nodes hosting non-reschedulable microservices, i.e. $\forall m \in M_a, a \in A \backslash \bar{A}$, are forced to be included as active virtual nodes, thus contributing to the overall deployment cost. Finally, constraint \eqref{eq_cons7} expresses the integer nature of the problem.

Intuitively, the resulting problem formulation can be reduced to a Bin Packing problem formulation when the rescheduling order and locality constraints are neglected, hence its complexity is NP-Hard \cite{martello1990lower}. For this reason, it is unpractical for large-scale deployment scenarios composed by a high number of microservices and virtual nodes. In the next section, we propose a heuristic algorithm approximating the optimal solution in order to lower the microservice rescheduling computational complexity.

\subsection{Heuristic re-orchestration algorithm}
 Essentially, the proposed optimal formulation solves to two tasks simultaneously: i) it computes a new deployment configuration $S_{\hat{A}}$ that, given the microservices locality requirements, minimizes the number of multi-cloud virtual nodes based on their price, ii) it computes a microservice rescheduling order that, given the initial deployment configuration $S_{A}$, ensures that $S_{\hat{A}}$ is achieved without service disruption. Following this observation, instead of performing these tasks at the same time, we approximate the optimal solution by designing a two-step heuristic algorithm in order to prioritize a significantly lower computational complexity while tolerating a minor increase in the probability of QoS violation and service disruption.

 \begin{algorithm}[t] 
\caption{Allocation phase}\label{alg:heuristic_phase_one}
\begin{algorithmic}[1]
\State \textbf{Input:} $\hat{A},N,p_n,\Delta C_n, \ell^{(n,n')}_{a,m,m'}, \hat{r}_{a,m}$
\State \textbf{Output:} $\tilde{S}_{\hat{A}} = \{ s_{a,m,n} \}$
\State Initialize $s_{a,m,n} = 0$,  $\forall m \in M_a, \forall a \in \hat{A}, \forall n \in N$ 
\State Initialize cluster resource occupation $r_{N_i} = 0$ 
\State Initialize node resource availability $\tilde{\Delta} C_n = \Delta C_n$ 
\For{$a \in \hat{A}$}
    \For{$i \in \{0,...,R-1\}$}
     %   \State Compute current cluster resource occupation \phantom{xxxxxxxx} $r_{R_i} =  \sum_{n \in R_i} \tilde{\Delta} C_n$ 
      %  \State Compute cluster score as $c_{R_i} = \frac{p_n r_{R_i}}{\Delta C_n}$
      \State Compute cluster score $c_{N_i} = \frac{\sum_{n \in N_i} \tilde{\Delta} C_n}{C_n}$
    \EndFor
    %\State Sort $c_{R_i}$ in decreasing order
    \State Select $\tilde{N}_i = \text{argmin}\{c_{N_i}\}$ 
        \State Compute $\!\tilde{M}_a \!=\! \{\!m \!:\! \ell^{(n,n')}_{a,m,m'} \!= \! 0 , m' \! \in M_a, n \!\in \tilde{N}_i, n' \! \in N_i \}$
        \For{$m \in \tilde{M}_a$}
           % \For{$n \in \tilde{R}_i$}
           %     \State Compute virtual node $c_{n} = \sum_{n \in R_i} s_{a,m,n} \hat{r}_{a,m}$
           % \EndFor
           % \State Sort $c_{n}$ in decreasing order 
           % \State Select node $\tilde{n}$ of maximum score such that $\hat{r}_{a,m} \leq \Delta C_n$
            \State Select $\tilde{n} = \text{argmin}\{\tilde{\Delta} C_n : n \in  \tilde{N}_i\ \!\land \hat{r}_{a,m} \leq \Delta C_n\}$ 
            \If{$\tilde{n} \neq \{\emptyset\}$}
                \State $s_{a,m,n} = 1$   
                \State $\tilde{\Delta} C_n \leftarrow \tilde{\Delta}C_n - \hat{r}_{a,m}$
            \EndIf
        \EndFor
\EndFor
\State Compute $\bar{M} = \{m :  m \in M_a \backslash \tilde{M}_a, \forall a \in \hat{A}  \}$ 
\State Sort $\bar{M}$  in decreasing order of $\hat{r}_{a,m}$ 
\For{$m \in \bar{M}$}
    \State Select $\tilde{n} = \text{argmin}\{p_n\tilde{\Delta} C_n : \hat{r}_{a,m} \leq \Delta \tilde{C}_n\}$ 
            \If{$\tilde{n} \neq \{\emptyset\}$}
                \State $s_{a,m,n} = 1$   
                \State $\tilde{\Delta} C_n \leftarrow \tilde{\Delta}C_n - \hat{r}_{a,m}$
            \EndIf
\EndFor   
\end{algorithmic}
\end{algorithm}

 \begin{algorithm}[t] 
\caption{Ordering phase}\label{alg:heuristic_phase_two}
\begin{algorithmic}[1]
\State \textbf{Input:} $S_A,\tilde{S}_{\hat{A}},\Delta C_n, \hat{r}_{a,m}, r_{a,m}$
\State \textbf{Output:} $x^{t}_{a,m,n}$ 
\State Initialize $x^{t}_{a,m,n} = 0$,  $\forall m \in M_a, \forall a \in \hat{A}, \forall n \in N, \forall t \in T$ 
\State Initialize $M = \{m :  m \in M_a, \forall a \in \hat{A}  \}$ 
%\State Compute $z^{n,n'}_{a,m} = \{z^{n,n'}_{a,m} = 1 : s_{a,m,n}=1 \land \tilde{s}_{a,m,n}=1$
\State Initialize scheduling slot $t = 0$
\State Initialize node resource availability $\tilde{\Delta} C_n = \Delta C_n$ 
%\State Compute microservice score \\ $c_{a,m,n} = \{ c_{a,m,n} : \frac{\hat{r}_{a,m}}{s_{a,m,n}\tilde{\Delta} C_n } \quad \text{if} \quad  s_{a,m,n}=1\}$
\For{$m \in M_a, \forall a \in \hat{A}$}
    \For{$n \in N$}
        \If{$s_{a,m,n}=1$}
            \State Compute score $c_{a,m,n} = \frac{\hat{r}_{a,m}}{s_{a,m,n}\tilde{\Delta} C_n}$
        \EndIf
    \EndFor
\EndFor
%\State Sort microservices in ascending order of $c_{a,m,n}$

\While{$\exists m \in M$} \label{heu2_first_while_step}
    %\State $T = \{ m : x^{t}_{a,m,n} = 0 \forall n \in N \}
    \State Select $m$ and $n$ s.t. $\text{argmin}\{c_{a,m,n}\}$
    \If{$\tilde{\Delta} C_n - \hat{r}_{a,m} \geq 0$}
        \State $x^{t}_{a,m,n} = 1$
        \State Update $\tilde{\Delta} C_n \leftarrow \tilde{\Delta}C_n - \tilde{s}_{a,m,n} \hat{r}_{a,m}$
        \State Update $\tilde{\Delta} C_{n'} \leftarrow \tilde{\Delta}C_{n'} + s_{a,m,n'} r_{a,m}$
        \State Remove $m$ from $M$ and update $c_{a,m,n}$ \label{heu2_last_while_step}
        \State Increase $t \leftarrow t+1$
    \Else 
        \State Select next $m'\neq m$ and $n$ s.t. $\text{argmin}\{c_{a,m,n}\}$
        \State Repeat steps \ref{heu2_first_while_step}-\ref{heu2_last_while_step}
        \If{$\nexists m$ reschedulable in time slot $t$}
            \State Select $m'$ s.t. $\text{argmax}\{r_{a,m}\}$
            \State Update $\tilde{\Delta} C_{n} \leftarrow \tilde{\Delta}C_{n} + s_{a,m',n} r_{a,m'}$
            \State Remove $m'$ from $M$ and update $c_{a,m,n}$
            \State Increase $t \leftarrow t+1$
        \EndIf
    \EndIf
\EndWhile
\end{algorithmic}
\end{algorithm}
 
  In the first phase, referred to as \textit{Allocation} phase, the algorithm ignores the rescheduling order as well as the initial deployment configuration $S_{A}$ and focuses on the computation of a deployment configuration $\tilde{S}_{\hat{A}}$ that minimizes the deployment cost while satisfying the locality constraints with high probability. We report the main steps of this phase in Algorithm \ref{alg:heuristic_phase_one}. In lines 6-19 we first progressively deploy microservices belonging to the same application that have some locality constraint. In detail, in lines 7-9 we assign to each cluster a score $c_{N_i}$ that reflects the percentage of overall available resources. Then, in lines 10-18 we select the cluster with the lowest resource occupation, and we allocate the microservices on the related least-expensive virtual nodes having the currently highest resource occupation as quantified by the metric $p_n\tilde{\Delta}C_{n}$. Successively, in lines 20-28 we sort the reaming microservices, defined by the set $\bar{M}$, in decreasing order of requested resources $\hat{r}_{a,m}$ and we progressively allocate them on the least-expensive and most-loaded virtual node employing the same metric used for the co-located microservices. 
  The proposed allocation procedure reduces the chances of locality constraints violations by evenly distributing co-located microservices across different clusters and, at the same time, it mitigates deployment cost by prioritizing the allocation on active nodes having a low price.

  %linwe minimize the chance to violate QoS constraints by selecting virtuals nodes of clusters with a low resource occupation, denoted by $c_{R_i}$,  allocating microservices on virtual cluster having a low resource consumption, denoted by $c_{R_i}$ across the various nodes cluster in order to balance the overall cluster resource occupation, which may prevent microservices co-location feasibility. Then, we deploy the remaining microservices according to a best-fit approach that sorts microservices  in decreasing order of resource request $\hat{r}_{a,m}$ and sequentially allocate them to best fitting virtual node. 
  In the second phase, referred to as \textit{Ordering} phase, the algorithm computes a feasible rescheduling order of microservices that reshapes the initial deployment configuration $S_{A}$ into the configuration $\tilde{S}_{\hat{A}}$, that was obtained in the \textit{Allocation} phase, by enforcing constraint \eqref{eq_cons3} in the optimal formulation. We report the main steps of this phase in Algorithm \ref{alg:heuristic_phase_two}. The general idea of the ordering scheme is to reschedule in each time slot the microservice having the minimum impact on the resource load of the allocated node so to lower the chance of an unfeasible rescheduling step (i.e., the allocation on a node that cannot accommodate the resources requested by the microservice in that time slot). In the case such event occurs, we turn off the microservice in order to free up the resources on the currently allocated node, thus easing the rescheduling of the remaining microservices in next time slots.  In detail, in lines 7-13 we compute the aforementioned score $c_{a,m,n}$ for each microservice. Then, as long as there are microservices to reschedule, in line 15 we select the candidate microservice according to the minimum score. In line 16 we check the allocation feasibility on the destination node, that is identified by $\tilde{s}_{a,m,n}$, in terms of resource availability.  If this is the case, in line 17-21 we perform the rescheduling in time slot $t$, we update the available resources on the source and destination nodes accordingly (the source node is the original allocation of the rescheduled microservice as indicated in $S_{\hat{A}}$), and we proceed with next scheduling slot $t+1$. Otherwise, in lines 23-24 we select the next microservice of minimum score and repeat the previously defined steps. In the scenario where no microservice can be rescheduled according to $\tilde{s}_{a,m,n}$ due to resource unavailability in rescheduling slot $t$, we remove the microservice currently requiring the highest amount of resources $r_{a,m}$ from the list of reschedulable microservices and we consider it as terminated by updating the node resource utilization on the related source node in lines 26-28. Then, we proceed with the next rescheduling slot in line 29.
  
  %In the case the algorithm fails to reschedules a microservice in the selected time slot due to resource unavailability on the destination virtual node, the non-reschedulable microservice is treated as temporarily turned off in order to complete the rescheduling process. The necessity to turnoff microservices, thus leading to service disruption, is a drawback derived from the two-step approach which trades a lower computational complexity for a penalty in term of possible service disruption compared to the optimal solution. Nonetheless, we mitigate this issue by designing a rescheduling ordering scheme that limits the number of turned off microservices.

\section{Performance evaluation}
\subsection{Simulation setup}

\begin{table}[t] 
\centering
\caption{Simulation parameters}\label{tab1}
\begin{tabular}{lc}
\toprule
Number of applications & \{10, 100\} \\
Microservices per application & 5 \\
Number of clusters & \{2, 4\} \\
Inter-cluster network latency & 50ms \\
Number of virtual nodes & \{16, 120\} \\
Number of vCPU & 8 cores \\
Available RAM & 64 GB \\
Normalized cost per multi-cloud node  & $p_n C_n = 1$ \\
Microservice CPU request & $[1.5,3.5]$ vCPU \\
Microservice RAM demand & $[0.2,0.5]$ GB\\
%Resource downscaling factor & $20\%$ \\
\bottomrule
\end{tabular}
\end{table}

\begin{figure*}[t]
     \centering
     \begin{subfigure}[b]{0.25\textwidth}
         \centering
         \includegraphics[width=\textwidth]{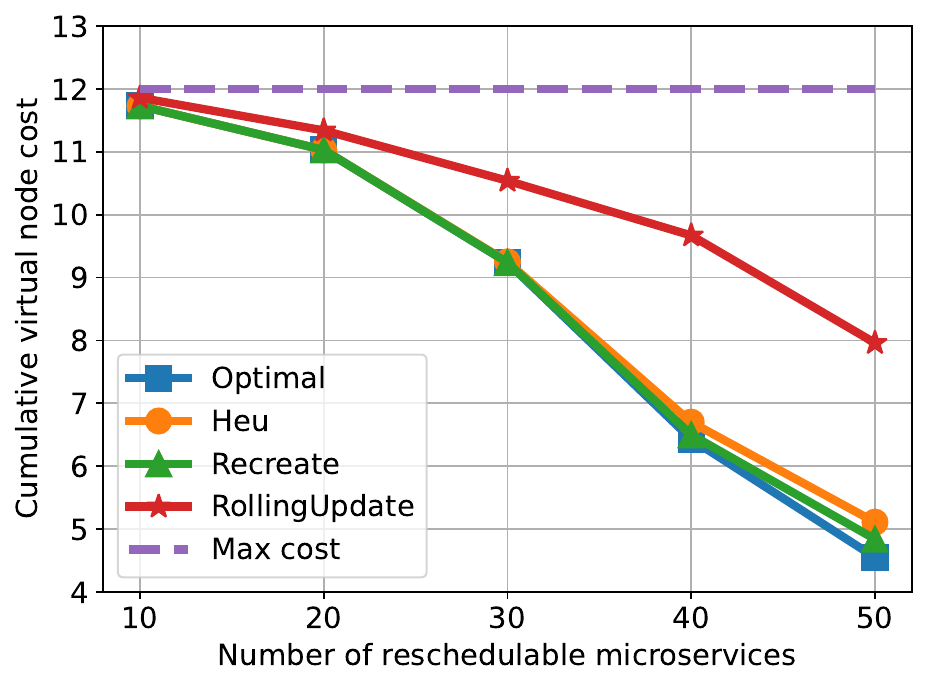}
         \caption{Deployment cost}
         \label{subfig:small_cost_single_region}
     \end{subfigure}
     \begin{subfigure}[b]{0.25\textwidth}
         \centering
         \includegraphics[width=\textwidth]{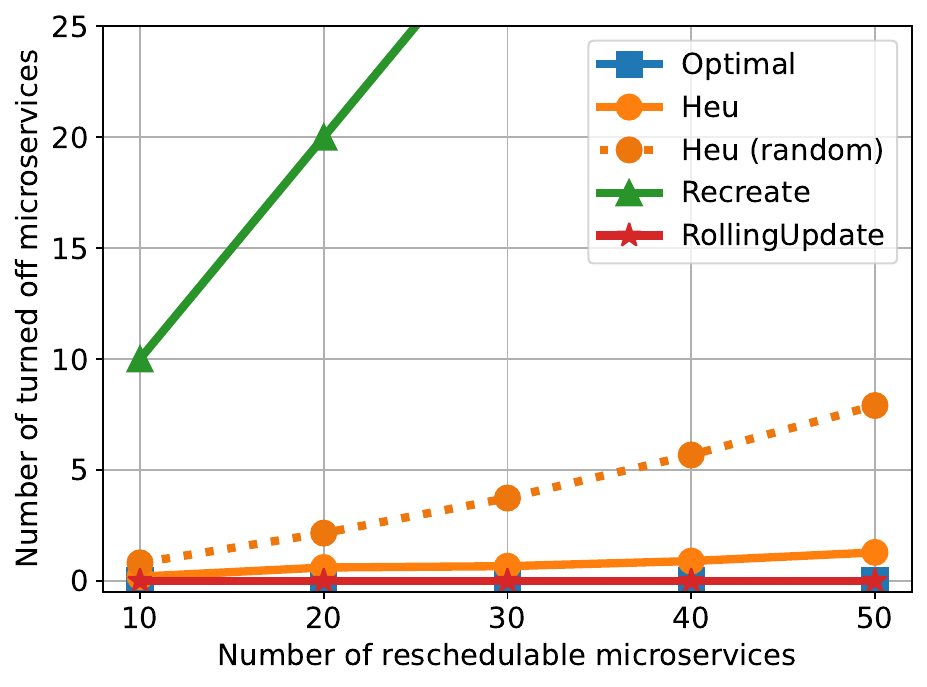}
         \caption{Disruption cost}
         \label{subfig:small_disruption_single_region}
     \end{subfigure}
     \begin{subfigure}[b]{0.25\textwidth}
         \centering
         \includegraphics[width=\textwidth]{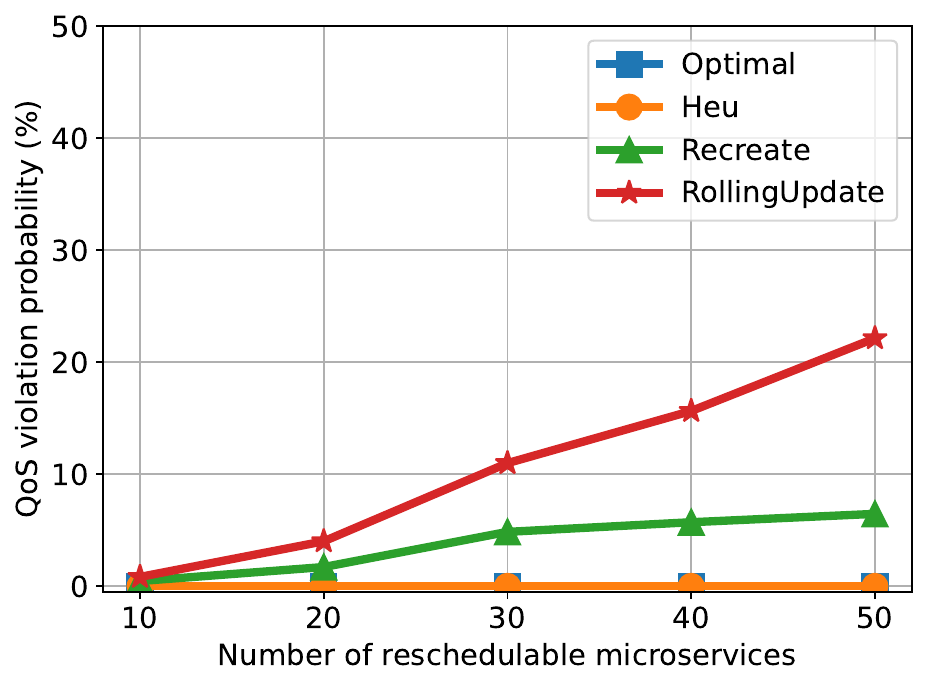}
         \caption{QoS violation}
         \label{subfig:small_qos_single_region}
     \end{subfigure}
        \caption{Results obtained when the number of regional cluster is $R=2$. The total number of virtual nodes in each cluster is $|N_0| = 4$, $|N_1| = 12$. The total number of microservices is 50 (10 apps of 5 ms each). The number of vertically down scaled microservices is 10.}
        \label{fig:small_deployment_single_region}
\end{figure*} 

\begin{figure*}[t]
     \centering
     \begin{subfigure}[b]{0.25\textwidth}
         \centering
         \includegraphics[width=\textwidth]{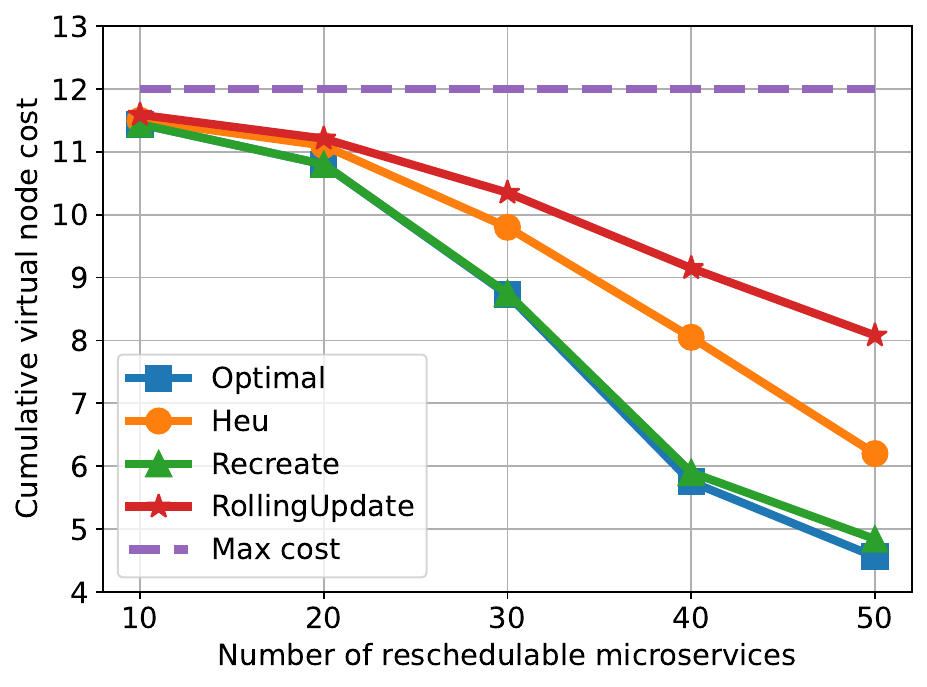}
         \caption{Deployment cost}
         \label{subfig:small_cost_multi_region}
     \end{subfigure}
     \begin{subfigure}[b]{0.25\textwidth}
         \centering
         \includegraphics[width=\textwidth]{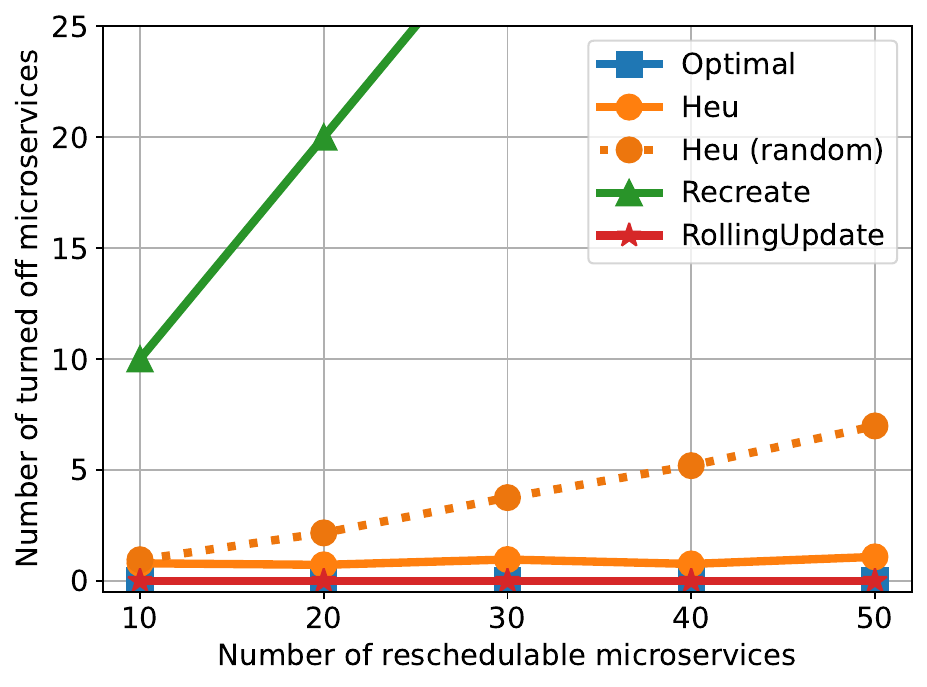}
         \caption{Disruption cost}
         \label{subfig:small_disruption_multi_region}
     \end{subfigure}
     \begin{subfigure}[b]{0.25\textwidth}
         \centering
         \includegraphics[width=\textwidth]{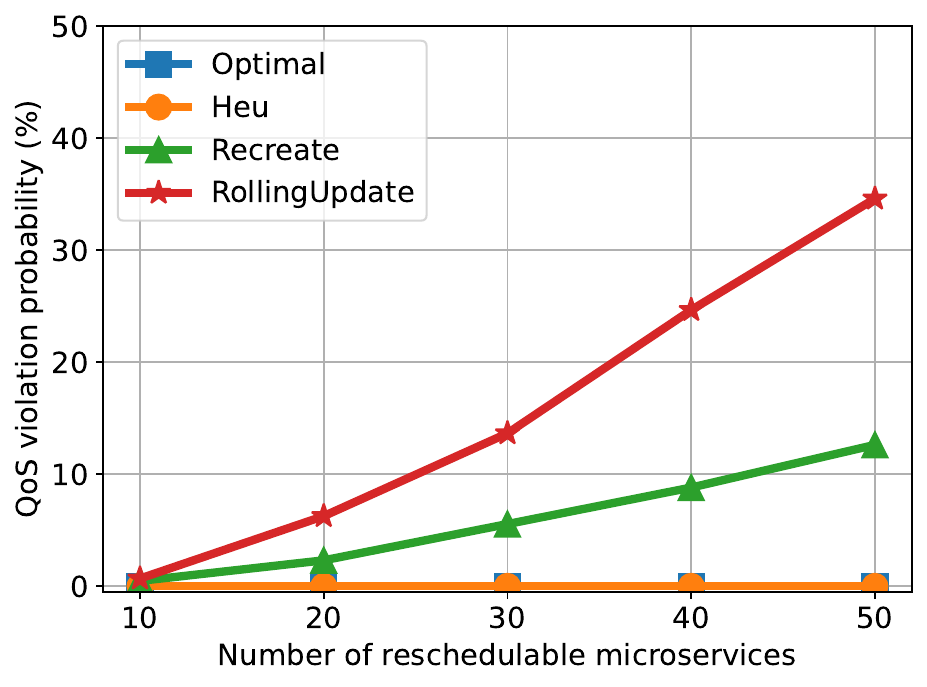}
         \caption{QoS violation}
         \label{subfig:small_qos_multi_region}
     \end{subfigure}
        \caption{Results obtained when the number of regional cluster is $R=4$. The total number of virtual nodes in each cluster is $|N_0| = |N_1| = |N_2| = |N_3| = 4$. The total number of microservices is 50 (10 apps of 5 ms each). The number of down scaled microservices is 10.}
        \label{fig:small_deployment_multi_region}
        \vspace{-0.5cm}
\end{figure*} 

We implemented the simulation environment using Python. In particular, we employed PySCIPOpt library to solve \eqref{eq_obj_func}-\eqref{eq_cons7}  using SCIP optimizer, which is a  non-commercial solver for mixed integer programming \cite{maher2016pyscipopt}. We compare the performance of the optimal and heuristic approaches against two benchmark schemes that simulate the same Kubernetes scheduling logic under two different microservice deployment procedures. We denoted these schemes as:  
\begin{itemize}
    \item \textbf{Rolling update}: it sequentially replaces reschedulable microservices by first allocating the updated instances before terminating the old ones.
    \item \textbf{Recreate}: it simultaneously replaces reschedulable microservices by first terminating all outdated instances before deploying the updated ones. 
\end{itemize}
Moreover, to provide a fair performance comparison with our approach, we replicated the Kubernetes scheduling decision process in order to minimize the deployment cost while preserving QoS requirements. Specifically, we combine the usage 
 of \textit{MostResourceFit}, \textit{NodeAffinity}, \textit{PodAffinity} scheduling plugins provided by the default Kubernetes scheduler implementation \cite{kube2023sched}. In particular, \textit{MostResourceFit} paired with \textit{NodeAffinity} allows prioritizing the packing of microservices on the cheapest heavy-loaded virtual nodes. Similarly, \textit{PodAffinity} allows prioritizing the intra-cluster co-location of delay-sensitive microservices by assigning high affinity values. 
We consider the following metrics  to compare  the performance of the aforementioned schemes:
\begin{itemize}
    \item \textbf{Deployment cost}: it measures the economic cost of the deployment after the completion of the re-orchestration procedure as the summation of the active virtual nodes prices.
    \item \textbf{Disruption cost}: it measures the service disruption severity as the number of turned-off microservices due to an unfeasible computation of a microservice rescheduling order.
    \item \textbf{QoS violation probability}: it measures the probability that a locality constraint is not satisfied after the completion of the re-orchestration procedure. It is computed as the share of violated locality constraints due to the allocation of microservices on different clusters.
\end{itemize}

We considered multiple simulation scenarios composed by a different number of applications and multi-cloud virtual nodes.
For each scenario, we computed the initial deployment configuration $S_A$ as follows. First, given a number of applications, we randomly generated a resource request configuration $r_{a,m}$ according to the values in \cite{detti2023mubench}, which analyzed the performance of several demos of microservice-based applications on different metrics including CPU utilization and memory consumption. Then, we assigned each microservice to the available nodes such that the average resource occupation of each node was at least $80\%$ of the maximum capacity (we considered the minimum value between CPU and RAM resources), which was assumed to be the same for every node. Then, we synthetically created the communication flows between microservices of the same application according to the Barabasi-Albert graph model \cite{zadorozhnyi2012structural}. The latter can be used to generate realistic topologies of communication flows between microservices following a preferential node attachment rule as analyzed in \cite{podolskiy2020weakest}. We associated to each flow a maximum tolerable latency by uniformly sampling the interval $[20ms,100ms]$, whose boundaries express latency-critical and latency-agnostic communication flows, respectively. Given these values, we generated the locality constraints $\ell^{(n,n')}_{m,m'}$ for each microservice pair given the inter-cluster network latency, which was assumed to be the same between each cluster.  This procedure provides a starting deployment configuration that always satisfies the locality constraints and that achieves a high resource utilization. 

Based on such deployment, we assumed that the $20\%$ of the deployed applications (randomly selected) required a resource down scaling in order to improve the resource utilization efficiency during low-intensity workload regimes. We arbitrarily selected this percentage in order to reproduce a conservative and more realistic scenario where only a fraction of the deployed applications requires a new resource reconfiguration.  We down scaled by $50\%$ the CPU resources assigned to the related microservices, hence the corresponding new CPU allocation was computed as $\hat{r}^{(cpu)}_{a,m} = 0.5 \times r^{(cpu)}_{a,m}$. Conversely, we considered as unchanged the memory occupation, hence $\hat{r}^{(ram)}_{a,m} =  r^{(ram)}_{a,m}$. We determined these scaling parameters based on the CPU and memory consumption statistics of applications deployed on the Alibaba Cloud cluster, as documented in \cite{everman2021improving}. The analysis showed a periodic pattern for the CPU usage, that fluctuated around the $50\%$ of its average value, and a relatively stable pattern for the memory occupation.

%We selected these scaling parameters based on the CPU and memory consumption statistics of the Alibaba Cloud cluster as reported in \cite{everman2021improving}, which showed a periodical pattern behavior for the CPU usage that fluctuated within the $50\%$  of its average value and  roughly constant pattern behavior for memory occupation.
%we down scaled by $50\%$ the CPU resources assigned to the $20\%$ of the deployed applications (randomly selected), hence for each related microservice the corresponding new CPU allocation was computed as $\hat{r}^{(cpu)}_{a,m} = 0.5 \times r^{(cpu)}_{a,m}$. We selected these parameters according to he overall CPU utilization and memory consumpation across 7 day on the Alibaba cloud cluster as reported in \cite{everman2021improving}, which showed that periodical fluctations of the resources across 7 days, where for the CPU was roughly in the range of 50 of the average value, whereas the memory showed very limited sensitivity to variations that for our simulations.

%and we compute the new deployment configuration according to the various considered schemes. %othe amount the of allocated reousoruce of the $30\%$, i.e. $\hat{r}_{a,m} = 0.3 * r_{a,m}$. %vertically scaling down the resources allocated to the related microservices as $\hat{r}_{a,m} = 0.3 * r_{a,m}$, where  $\rho$ is referred as downscaling factor (note that the same reasoning applies if upscaling was considered). 
We evaluated the re-orchestration performance of the considered schemes by repeating the aforementioned process 50 items and averaging the results. We resume the main simulation parameters in Table \ref{tab1}.

\subsection{Simulation results}

\begin{figure*}[th]
     \centering
     \begin{subfigure}[b]{0.3\textwidth}
         \centering
         \includegraphics[width=\textwidth]{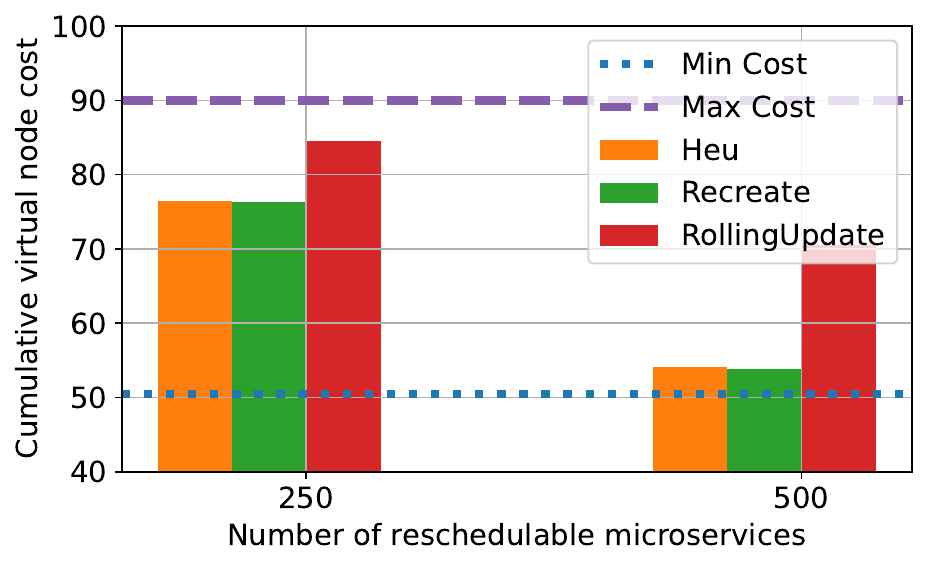}
         \caption{Deployment cost}
         \label{subfig:large_cost_multi_region}
     \end{subfigure}
     \begin{subfigure}[b]{0.3\textwidth}
         \centering
         \includegraphics[width=\textwidth]{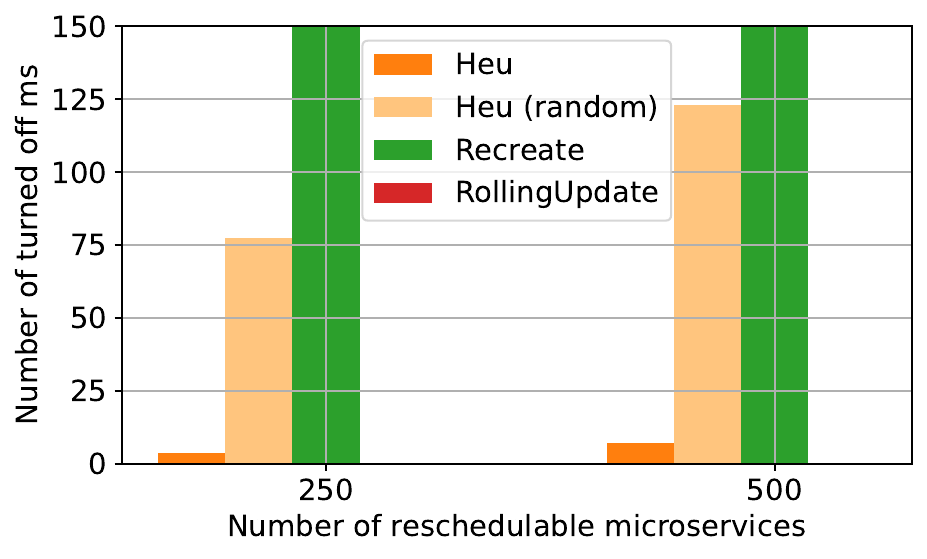}
         \caption{Disruption cost}
         \label{subfig:large_disruption_multi_region}
     \end{subfigure}
     \begin{subfigure}[b]{0.3\textwidth}
         \centering
         \includegraphics[width=\textwidth]{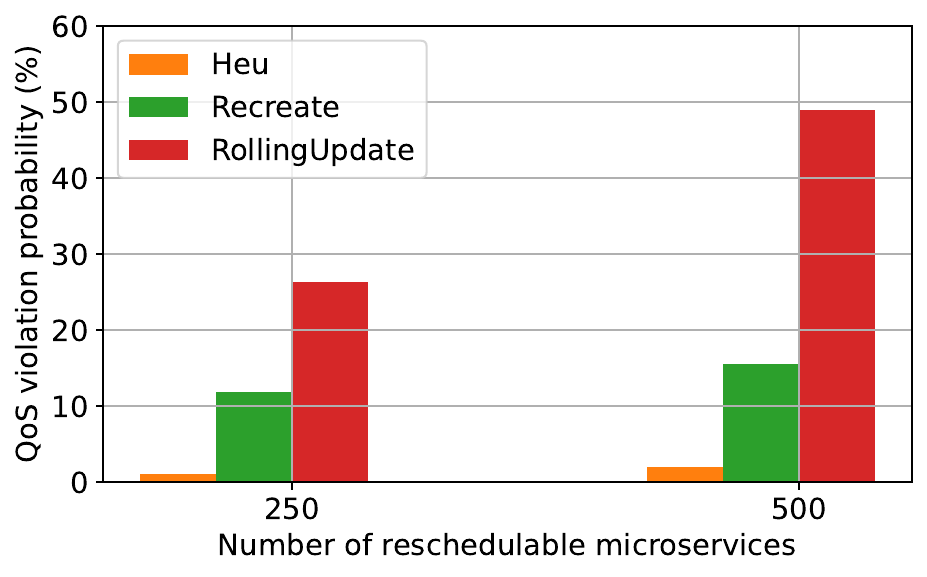}
         \caption{QoS violation}
         \label{subfig:large_qos_multi_region}
     \end{subfigure}
        \caption{Results obtained when the number of regional cluster is $R=4$. The total number of virtual nodes in each cluster is $|N_0| = |N_1| = |N_2| = |N_3| = 30$. The total number of microservices is 500 (100 apps of 5 ms each). The number of vertically down scaled microservices is 100.}
        \label{fig:large_deployment_multi_region}
        \vspace{-0.5cm}
\end{figure*}

\begin{figure}[th]
    \centering
     \includegraphics[width=0.3\textwidth]{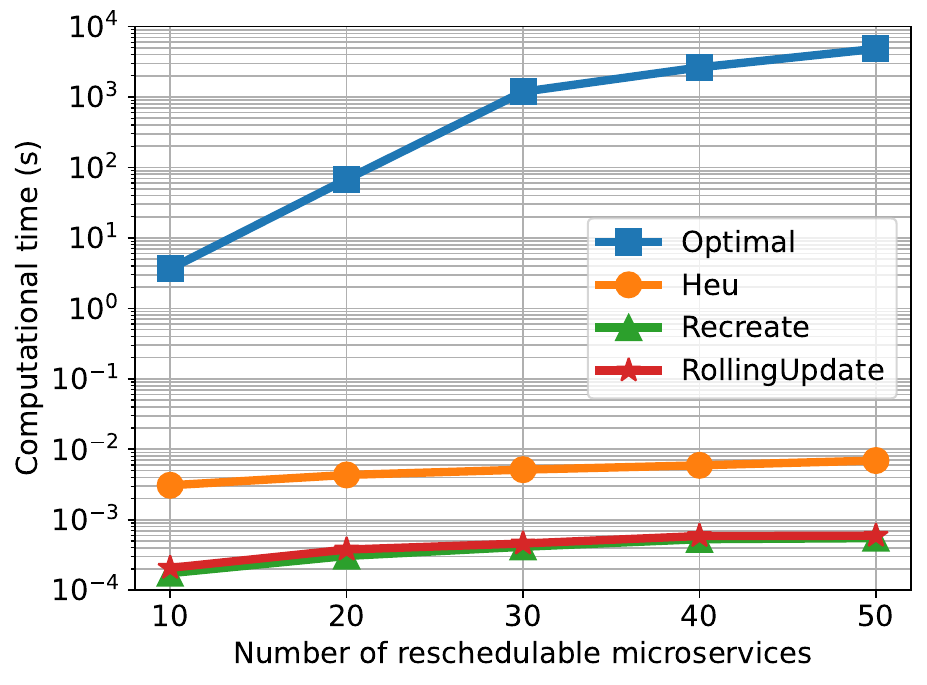}
     \caption{Solution computation time when the number of regional cluster is $R=4$ and the number of microservices is 50.}
     \label{fig:time_complexity}
     \vspace{-0.5cm}
\end{figure}

 We initially consider a small deployment configuration due to the time required by the optimal formulation to compute the solution. In this way, we can assess the quality of the heuristic solution approximation while analyzing the behavior of other schemes. In Fig. \ref{fig:small_deployment_single_region}, we show the results obtained for each considered metric when $R =2$. 
 In general, in Fig. \ref{subfig:small_cost_single_region}, we observe that all schemes provide better cost gains as the number of reschedulable microservices increases compared to the initial deployment cost indicated by Max Cost curve. A higher number of reschedulable microservices ensures a higher degree of freedom to each algorithm, which can compute a more effective microservice repacking that leads to a lower number of active virtual nodes. In detail, the Rolling Update scheme provides the worst cost minimization performance due to the fact that it progressively computes the assignment of each microservice according to the current load in each virtual node that keeps varying during the rescheduling phase. This effect severely degrades its performance since the rescheduling decision performed in a given time slot may be largely suboptimal in future time slots.
 
Conversely, the heuristic and Recreate scheme provides a good approximation of the optimal solution as they both  compute the rescheduling order as if it was meant for an initial deployment phase, where applications are deployed for the first time and virtual nodes are empty.  This strategy guarantees that the resource load in each virtual node monotonically increases after each rescheduling step, ensuring a reliable resource utilization efficiency.

However, compared to the Recreate scheme which requires the complete shutdown of the deployed applications before computing the microservice rescheduling, the proposed heuristic ensures a close-to-zero disruption cost like the optimal algorithm and Rolling Update schemes (which are disruption-free by design) as shown in Fig. \ref{subfig:small_disruption_single_region}. This is due to the Ordering phase that computes a non-trivial rescheduling order capable of accommodating most of the temporary burst of resource occupation derived from the  simultaneous coexistence of the old and updated microservice instances. We better highlight the benefit of this approach by also showing the disruption cost obtained by a heuristic algorithm using a random Ordering phase, denoted as Heu (Random), where the rescheduling order is indeed randomly generated. 

Similarly, as shown in Fig. \ref{subfig:small_qos_single_region}, the heuristic and optimal schemes also ensure a zero probability of violating the locality constraints thanks to a suitable deployment of microservices in both clusters. The conservative approach implemented by the Allocation phase of the heuristic scheme ensures that microservices requiring co-location are reassigned first and are evenly distributed among the available clusters. This strategy reduces the chances of scenarios where the aggregated capacity of a cluster is supposed to accommodate co-located microservices, but the subsequent node assignment fails due to an insufficient resource availability. In contrast, the Recreate and Rolling Update schemes provide considerable higher QoS violation probability, which increases as more microservices are rescheduled. This is due to the fact that they compute the microservice rescheduling based on a greedy approach that jointly consider the cost minimization together with the QoS fulfillment. This behavior may lead to conflicting rescheduling decisions as the cost minimization tends to compact the microservice deployment thanks to the metrics \textit{MostResourceFit} and \textit{NodeAffinity} on least expensive nodes, saturating them and thus increasing the chances to fail the allocation of microservices with \textit{PodAffinity} requirements.

To assess performance in a more complex multi-cloud scenario, in Fig. \ref{fig:small_deployment_multi_region} we also plot the results when the number of multi-cloud cluster is $R=4$ while maintaining the same number of virtual nodes as in the previous scenario (in other words, each region cluster has a lower capacity). In general, we observe a degradation of performance of all schemes due to the more challenging locality constraints. In particular, as shown in Fig. \ref{subfig:small_cost_multi_region}, the optimal solution suffers the least performance loss as it is capable of adapting the microservice rescheduling order to guarantee no disruption and QoS violation with a minimum loss in terms of deployment cost saving. Similarly, even though with a more visible performance gap, the heuristic replicates the optimal scheme dynamic and trades off a bigger deployment cost to maintain a low disruption cost and QoS violation probability, as depicted in Fig. \ref{subfig:small_disruption_multi_region} and in Fig. \ref{subfig:small_qos_multi_region}, respectively. Conversely, the growing complexity in the accommodation of the locality constraints further amplifies the rescheduling procedure limits of the Recreate and Rolling Update schemes, as previously discussed. As a matter of fact, in a real-world scenario,  the resulting re-orchestrated deployment configuration may be unpractical due to the high QoS violation probability. 

In Fig. \ref{fig:large_deployment_multi_region}, we consider a larger deployment configuration to better analyze the heuristic scheme performance in a more realistic scenario composed by a higher number of applications and nodes. Note that we omit the evaluation of the optimal scheme due to the prohibitive computational time. Moreover, since all the considerations previously outlined for each scheme still hold, we restrict the analysis by showing the results when $50\%$ and $100\%$ of microservices can be rescheduled. In general, the heuristic scheme provides a considerable higher gain in terms of deployment cost minimization compared to the Rolling Update approach as shown in Fig. \ref{subfig:large_cost_multi_region}. In particular, the achieved cost is close to the minimum obtainable value, indicated by Min Cost curve, that we plot as performance lower bound in order to better highlight the effectiveness of the proposed scheme. Similarly, we observe a slight increase in the disruption cost and QoS violation probability achieved by the heuristic scheme that nonetheless outperforms the Recreate and Rolling Update approaches as shown in Fig.\ref{subfig:large_disruption_multi_region} and in Fig.\ref{subfig:large_qos_multi_region}, respectively. We remark the advantage provided by the Ordering phase of the proposed heuristic over a random microservice ordering. This is due to the fact that the Ordering phase prioritizes a conservative approach that first reschedules microservices having a low impact on the overall resource occupation of the destination node. As a result, this strategy provides a higher chance to successfully accommodate the rescheduled microservices in subsequent time slots and help containing the disruption cost when the number of reschedulable microservices increases.

Finally, we report the computational performance in Fig.\ref{fig:time_complexity}. As already anticipated, the optimal solution is unsuited for large-scale deployment scenarios due to the high computational complexity that is required even for small deployments. Differently, the proposed heuristic algorithm  significantly reduces the calculation time and serves as a practical alternative to the optimal approach. Moreover, although the Recreate and Rolling Update schemes are characterized by faster computation time, the substantial performance improvements provided by the heuristic scheme makes the latter a superior choice for real-world scenarios.

\section{Conclusion}
We addressed the problem of lowering the deployment cost of microservice-based applications hosted on  geographically distributed multi-cloud systems throughout a suitable microservice re-orchestration scheme. Given an initial deployment configuration, the proposed optimization-based strategy efficiently repacks microservices at runtime on the available virtual nodes based on a rolling update scheduling logic. The resulting microservice reallocation minimizes the economic cost associated to the usage of multi-cloud virtual nodes and, at the same, preserves the locality requirements of microservices by guaranteeing their co-location on virtual nodes belonging to the same cluster. To overcome the prohibitive computational complexity of the optimal approach, we also proposed a low-complexity two-phase heuristic to approximate the optimal solution. The results showed that the proposed approach outperformed baseline schemes, leveraging Kubernetes scheduler configuration options, in terms of cost deployment minimization, disruption cost and QoS violation.

\section*{Acknowledgements}
This work has received funding from the EU Horizon Europe R\&I Programme under Grant Agreement no. 101070473 (FLUIDOS).

\bibliographystyle{IEEEtran}
\bibliography{bibliography}

\end{document}